\documentclass[12pt,preprint]{aastex}
\usepackage{times}

\newcommand{\stereo}{{\it STEREO}}
\newcommand{\sdo}{{\it SDO}}

\shorttitle{Imaging Far-Side Sun}
\shortauthors{Zhao et al.}

\begin{document}
\title{Imaging the Sun's Far-Side Active Regions by Applying Multiple 
Measurement Schemes on Multi-Skip Acoustic Waves}

\author{Junwei Zhao\altaffilmark{1}, Dominick Hing\altaffilmark{2}, Ruizhu 
Chen\altaffilmark{1}, and Shea Hess Webber\altaffilmark{1}}
\altaffiltext{1}{W.~W.~Hansen Experimental Physics Laboratory, Stanford
University, Stanford, CA 94305-4085, USA}
\altaffiltext{2}{Department of Physics, Stanford University, Stanford,
CA 94305-4060, USA}

\begin{abstract} 
Being able to image active regions on the Sun's far side is useful for modeling 
the global-scale magnetic field around the Sun, and for predicting the arrival 
of major active regions that rotate around the limb onto the near side.
Helioseismic methods have already been developed to image the Sun's far-side 
active regions using near-side high-cadence Doppler-velocity observations; 
however, the existing methods primarily explore the 3-, 4-, and 5-skip 
helioseismic waves, leaving room for further improvement in the imaging quality 
by including waves with more multi-skip waves. Taking advantage of the facts
that 6-skip waves have the same target-annuli geometry as 3- and 4-skip waves, 
and that 8-skip waves have the same target-annuli geometry as 4-skip waves, 
we further develop a time--distance helioseismic code to include a total 
of 14 sets of measurement schemes. We then apply the new code on the 
\sdo/HMI-observed Dopplergrams, and find that the new code provides substantial
improvements over the existing codes in mapping newly-emerged active 
regions and active regions near both far-side limbs. Comparing 3 months of 
far-side helioseismic images with the \stereo/EUVI-observed 304~\AA\ images,
we find that 97.3\% of the helioseismically detected far-side active regions 
that are larger than a certain size correspond to an observed region with 
strong EUV brightening. The high reliability of the new imaging tool will 
potentially allow us to further calibrate the far-side helioseismic images 
into maps of magnetic flux.
\end{abstract}


\section{Introduction}
\label{sec1} 

The bulk of the Sun is not transparent to electromagnetic waves, thus active 
regions (ARs) on the Sun's far side are not directly visible to us; however,
the Sun's helioseismic waves, mostly $p$-mode (or acoustic) waves, can propagate 
through the Sun's interior and bring far-side information to the near side. 
Most large far-side ARs are detectable through an analysis of such acoustic 
waves observed on the near side, and this allows us to monitor the evolution of 
large ARs throughout their lifetimes and to forecast arrivals of large ARs 
that rotate onto the Sun's near side from its far side. 

\citet{lin00} first detected a large AR on the Sun's far side by employing 
helioseismic holography method \citep{lin00b} and using the Doppler-velocity 
data observed by the {\it Solar and Heliospheric Observatory} / Michelson 
Doppler Imager \citep[{\it SOHO}/MDI;][]{sch95}. The far-side images from 
this first attempt, which utilized acoustic waves reflected twice on either 
side of the mapped far-side locations (2$\times$2-scheme hereafter), are 
limited to only the central area within about $50\degr$ (here, $\degr$ 
represents heliographic degree) from the Sun's far-side disk center. This 
limitation was later overcome by introducing a method that utilizes waves 
that are reflected once on one side and three times on the other side of 
far-side locations (1$\times$3-scheme hereafter), and the entire far side was 
thus able to be mapped \citep{bra01}. These helioseismic far-side maps, 
characterized by the phase shifts of helioseismic waves, are well correlated 
with the surface magnetic field and have a good potential to be calibrated 
into maps of unsigned magnetic flux \citep{gon07}.

\begin{figure}[!ht]
\epsscale{0.75}
\plotone{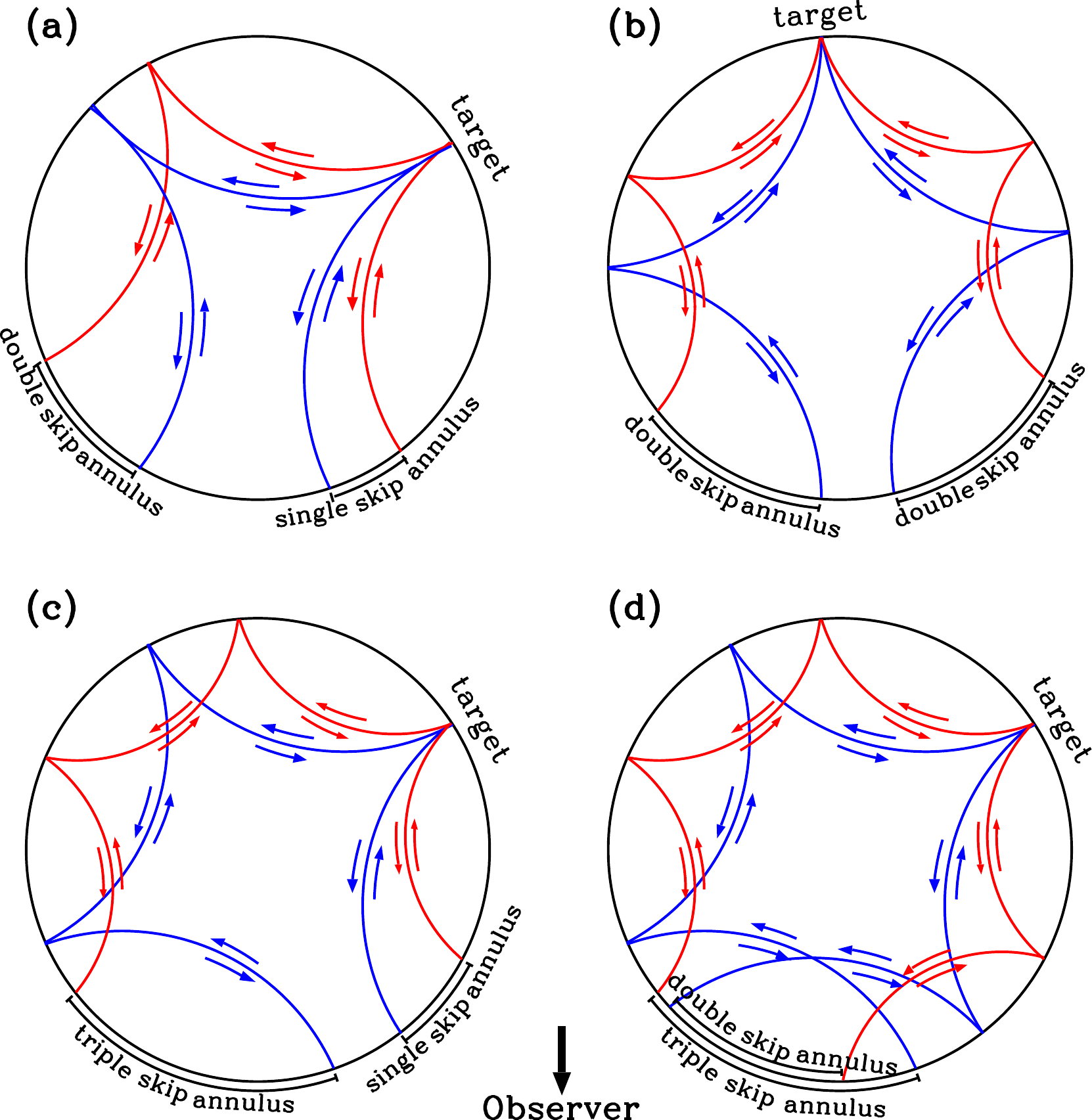}
\caption{Measurement schemes, viewed as great-circle slices through far-side 
target points, for (a) 1$\times$2-scheme for the sole branch of 3-skip waves; 
(b) 2$\times$2-scheme for the lower branch of 4-skip waves; (c) 1$\times$3-scheme 
for the lower branch of 4-skip waves; and (d) 2$\times$3-scheme for the lower 
branch of 5-skip waves. With the observer situated toward the bottom of the page, 
the lower half of each circle represents the near side of the Sun, and the 
upper half the far side. In each scheme, red curves indicate the ray paths 
for the waves with the shortest travel distance that is used in the corresponding 
measurement scheme, and blue curves for the waves with the longest travel 
distance. The ``target'' marked on the far side represents a sample location to 
be mapped, and the annuli marked on the near side indicate the ranges, within
which acoustic signals are taken for calculations. }
\label{f2}
\end{figure}

Time--distance helioseismology \citep{duv93}, another local helioseismic 
method, was also used to detect far-side ARs \citep{duv00}. Later, 
\citet{zha07} more systematically developed this method to include both 4-skip 
(composed of both the 2$\times$2- and 1$\times$3-schemes) and 5-skip 
(2$\times$3-scheme) acoustic signals, and the combination of both 4- and 5-skip 
waves helps to enhance the signal-to-noise ratio of the far-side images. 
\citet{ilo09} demonstrated that 3-skip acoustic signals (1$\times$2-scheme) 
can also be used in far-side imaging, the quality of which can be further 
improved by combining maps made from all of the 3-, 4-, and 5-skip helioseismic 
waves. All the above-mentioned measurement schemes are illustrated in 
Figure~\ref{f2}. Here, it is useful to caution that although both the helioseismic
holography and time-distance methods employ 2$\times$2- and 1$\times$3-schemes,
these two methods are fundamentally different in the measuring processes.

Prior to the availability of direct far-side observations, the reliability 
of imaging far-side ARs could only be assessed through applying the technique 
on numerical simulation data \citep{har08}, or comparing the detected 
far-side ARs with the near-side ARs before their rotation onto the far side or 
after their rotation onto the near side. More recently, the validity of 
helioseismic far-side imaging can be better evaluated by comparing with
the Sun's far-side EUV observations of the Extreme UltraViolet Imager 
(EUVI) onboard the twin {\it Solar TErrestrial RElations Observatory} (\stereo)
spacecraft \citep{how08}. In a series of articles, \citet{lie12, lie14, lie17} 
systematically compared \stereo/EUVI 304~\AA\ images, in which ARs 
exhibit as regions of enhanced emission, with helioseismic holography far-side 
images made from observations of both {\it Global Oscillations Network Group} 
\citep[GONG;][]{har96} and {\it Solar Dynamics Observatory} / Helioseismic and 
Magnetic Imager \citep[\sdo/HMI;][]{sch12a, sch12b}. They found that 95\% 
of the helioseismically-detected far-side ARs correspond to an observed
EUV brightening area, but only about 50\% of EUV brightening areas correspond 
to a helioseismically detected AR. This gives us a sense of the reliability 
and limitation of the helioseismic holography far-side imaging method.

Imaging the Sun's far-side ARs is useful for better space weather forecasting. 
Modeling coronal magnetic field and solar wind requires an input of global 
photospheric magnetic field, not just of the near side; however, the far-side 
magnetic field that is used in most models typically relies on observations 
of days earlier, or calculations from flux-transport models 
\citep[e.g.,][]{sch03, upt14}. Unfortunately, flux-transport models cannot 
model the ARs that continue to grow or that newly emerge on the far side, and
this would cause inaccuracies in the far-side magnetic-field data used in 
the global field and solar wind modeling. The global magnetic field with updated 
far-side AR evolution, particularly for ARs that rapidly emerge and grow on 
the far side, is important for a more accurate modeling: e.g., \citet{arg13} 
applied their ADAPT model \citep[Air-Force Data Assimilative Photospheric Flux 
Transport model;][]{arg10} on an updated magnetic field map with a 
newly emerged far-side AR detected by the helioseismic imaging method, 
and found that their modeled solar wind result is in better agreement with 
the observation than the result without including the new far-side AR.

However, the reliability and quality of helioseismic far-side images still
need to be improved for use in the forecasting models, and in fact can be 
improved. All the previous methods only explored the 3-, 4-, and 5-skip 
helioseismic waves, while waves with more skips are also useful but have not 
been explored in far-side imaging. In this article, taking advantage of the 
fact that 6-skip acoustic waves share the same surface target-annuli geometry 
with 3- and 4-skip waves, and that 8-skip waves share the same surface 
target-annuli geometry
with 4-skip waves, we further implement the 6- and 8-skip waves in the 
existing time--distance far-side imaging codes and develop a new code. 
Through comparing the new far-side helioseismic images with images from 
the holography method, as well as with \stereo's EUV 304~\AA\ observations, 
we assess the reliability and the AR-detection success rate of the new far-side 
imaging code. The article is organized as follows: we introduce our new 
method development in Section~\ref{sec2}, present new far-side imaging results 
in Section~\ref{sec3}, and assess the reliability and success rate of the 
new imaging method in Sections~\ref{sec4} and \ref{sec5}, respectively. 
We then discuss our results and give conclusions in Section~\ref{sec6}.

\section{Method}
\label{sec2}

\subsection{Multi-Skip Waves}
\label{sec21}

\begin{figure}[!ht]
\epsscale{0.70}
\plotone{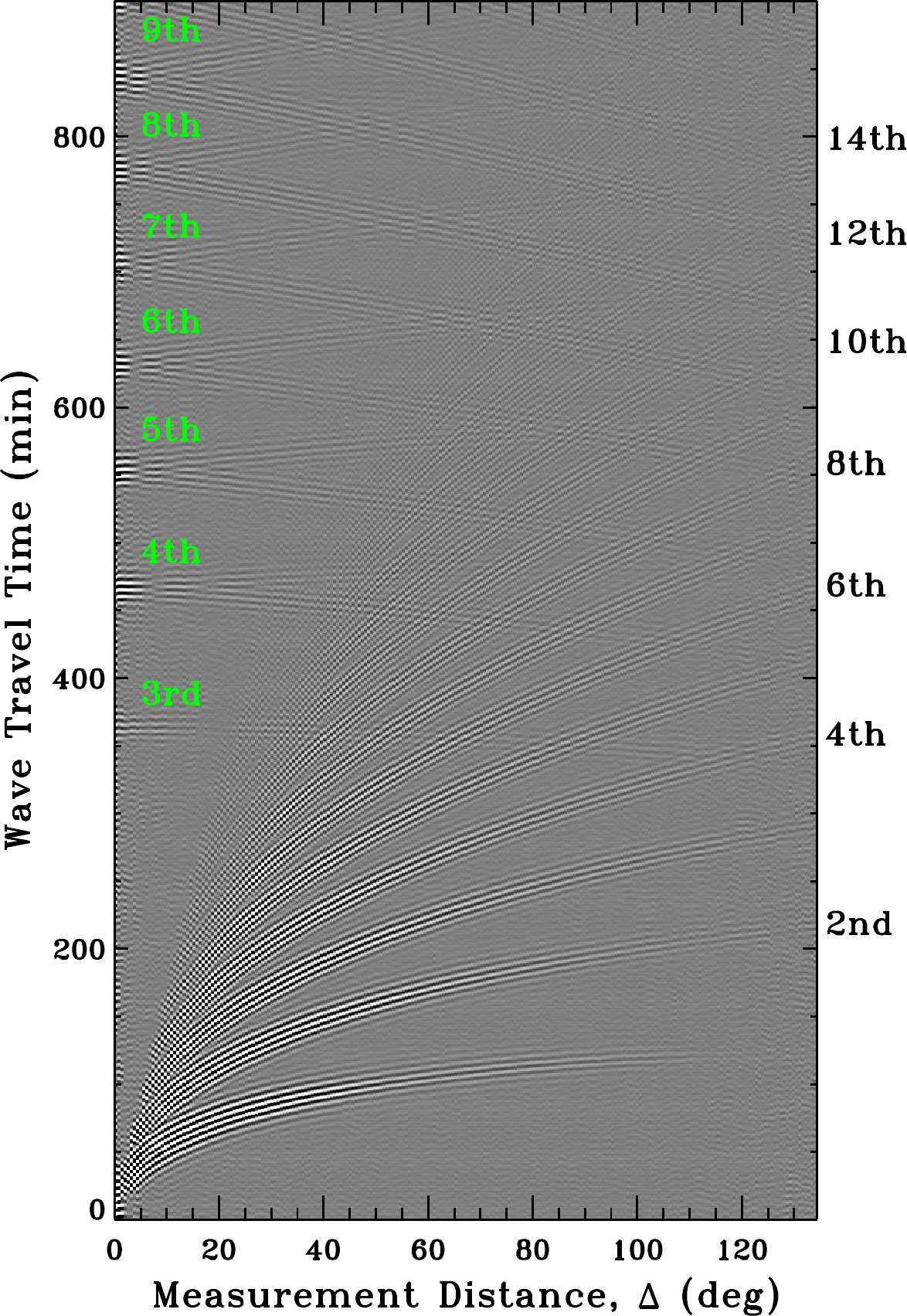}
\caption{Time--distance diagram obtained from cross-correlating 31 hours of 
\sdo/HMI Doppler-velocity data and displayed as a function of time lag 
(corresponding to wave travel time) and great-circle distance $\Delta$
between the pairs used for the cross-correlation calculation. On the right-hand 
side of the axis mark the $n$-th skip of acoustic waves traveling within the 
near side of the Sun. Green letters near the left axis mark the $n$-th skip 
waves that arrive back to the near side after traveling to the far side and 
experiencing $n$ surface reflections. }
\label{f1}
\end{figure}

As introduced in Section~\ref{sec1}, the previous time--distance far-side imaging 
method used 3-, 4-, and 5-skip acoustic waves \citep{zha07, ilo09}. Starting 
from the near side, these waves experience a total of 3, 4, and 5 
surface reflections during their travel to the far side and then back to 
the near side through the Sun's interior. However, as shown in Figure~\ref{f1}, 
in the time--distance diagram obtained from 31-hr Doppler observations of \sdo/HMI, 
not only the 3-, 4-, and 5-skip waves show clear signals, but the 6-, 7-, 8-, and 
9-skip waves are also visible within the duration of 900-min wave travel time. 
The time--distance diagram is made through cross-correlating oscillatory signals
between different surface locations, and is displayed as a function of time
lag and great-circle distance, $\Delta$,  between these surface locations, 
where the time lag in the cross-correlation functions is interpreted as travel 
time of acoustic waves from one surface location to the other. As can be seen 
in Figure~\ref{f1}, the correlation coefficients corresponding to the 6--9-skip 
waves, which is an indicator to the waves' oscillatory amplitude, is not much 
lower than those corresponding to the 3--5-skip waves, demonstrating that 
waves experiencing more surface reflections do not decay much more and are 
suitable to be included in the far-side imaging computations. The waves 
experiencing more than 9 skips may still be strong and clean, but their longer 
travel times make them less practical to be used in the far-side imaging method, 
which typically uses observational data of 1 day or slightly longer. 

The cross-correlation functions representing the 4- through the 9-skip waves 
(see Figure~\ref{f1}) exhibit a ``$<$" shape with its left-most turning-point 
at the measurement distance $\Delta = 0\degr$. Actually, the $\Delta = 0\degr$ 
for these signals that return from the far side represents a total of $360\degr$ 
travel distance around the globe, the lower branch of the ``$<$" shape represents 
a travel distance of $360\degr - \Delta$, and the upper branch represents a 
travel distance of $360\degr + \Delta$. The waves in the upper branches have 
a slightly faster travel speed than those in the lower branches of the same skip, 
and travel a longer distance while experiencing a same number of surface 
reflections. They correspond to the oscillation modes with a lower 
harmonic degree $\ell$. The time--distance diagram in Figure~\ref{f1} does 
not distinctly show the upper branch for the 3-skip waves, nor either branch 
of the 2-skip waves coming back from the far side, implying that these waves 
correspond to the low-$\ell$ modes that are either unresolvable in the \sdo/HMI 
observations or do not possess a sufficiently high signal-to-noise ratio to show 
up in this 31-hr observation. It is worth pointing out that multi-skip acoustic 
waves traveling within the near side of the Sun can also be seen in Figure~\ref{f1}, 
but these waves are not useful in imaging the Sun's far side. Shown as 
monotonically-upward ridges on the right side of the diagram, these waves can also
be long-lasting, and many of them survive more than 20 surface reflections before 
completely decaying away.

\citet{zha07} explored the lower branch of 4-skip waves and both branches 
of 5-skip waves, while \citet{ilo09} investigated the sole branch of 3-skip 
waves. This work further expands their analyses and includes the upper 
branch of 4-skip waves and both branches of the 6- and 8-skip waves in the 
new analysis method. This expansion takes advantage of the fact that measurements
of 6-skip waves using a 2$\times$4-scheme share the same surface target-annuli 
geometry as the 1$\times$2-scheme for 3-skip waves, measurements of 6-skip waves
using a 3$\times$3-scheme share the same surface geometry as the 2$\times$2-scheme 
for 4-skip waves, and measurements of 8-skip waves using 2$\times$6- and 
4$\times$4-schemes share the same target-annuli geometries as the 1$\times$3- and 
2$\times$2-schemes for 4-skip waves (see Figure~\ref{f2}). Although both 7- and 
9-skip signals are expected to be useful in the far-side imaging, they are 
not included in our new code. No existing measurement scheme exists to 
accommodate the 7-skip waves; although the 9-skip waves can be used in 
a 3$\times$6-scheme, same in surface geometry as the 1$\times$2-scheme for 3-skip 
waves, the low imaging quality from the upper branch of 8-skip
waves (see Sec.~\ref{sec32}) demonstrates that the final far-side imaging 
quality may not benefit much from including the 9-skip waves.

\subsection{Measurement Schemes}
\label{sec22}

The basic method for imaging the Sun's far-side ARs is described as follows. 
An acoustic wave starting from a near-side location expands like water 
ripples and seemingly sweeps across the whole Sun. Although the acoustic waves 
look much like surface ripples, they actually form so when acoustic waves 
travel into the interior and get refracted back continually from different 
depths of the solar interior. These ripples form the 1st-skip wave. The 
2nd-, 3rd-, 4th-, ..., skip waves follow when the 1st-skip waves are reflected 
back into the interior and refracted again, and thus form a series of 
expanding ripples. Depending on their modes, some waves can travel very far 
and continue back to the original near-side location and even beyond after 
a few surface reflections, shown as left-side signals in Figure~\ref{f1}. The 
waves that get reflected inside a far-side AR experience a reduction in their 
total travel times, either because the wave speed is relatively faster beneath 
sunspots \citep{kos00, har08}, or due to plages around ARs \citep{lin17}.
Such a travel-time reduction can be measured using near-side observations 
to show the location, size, and perhaps magnetic-field strength of the far-side 
AR.

Because of significant fluctuations and uncertainties in the travel times 
measured from the waves traveling around the globe, it is unrealistic to obtain
far-side AR information from measurements using merely one wave starting from 
one near-side location and ending at another near-side location. Instead, it 
is necessary to use all near-side locations that have a same 
great-circle distance (hereafter, distance means great-circle distance 
unless specified otherwise) to or from the far-side target, so that this 
far-side target effectively becomes a focus point where a great number of waves get
their surface reflections. Those near-side surface locations, one set as wave 
emitters and the other set as wave receivers, are pre-determined in order 
to save computational time in our calculation of the far-side images. In 
practice, a range of different travel distances are used for one measurement 
geometry. These locations form either an annulus or an arc segment in the 
near side of the Sun. Using 3-, 4-, and 5-skip waves as examples, 
Figure~\ref{f2} displays our measurement geometries illustrated in a great-circle 
slice through a far-side target. Measurements using 6- and 8-skip 
waves have the same surface target-annuli geometries but different interior 
paths. Table~\ref{tb1} lists measurement parameters for all the 14 measurement 
schemes to be used in this work. All the numbers listed in the table are
based on $0\degr$ solar $B$-angle. As a matter of fact, for the efficiency
and simplicity of the computations, all the images are calculated for the far 
side of the apparent solar disk without considering the $B$-angle variations.
The Carrington longitude and $B$-angle are only considered when remapping 
the far-side images into synchronic maps as shown in Sec.~\ref{sec4}.

\begin{table}[!t]
\begin{center}
\caption{Parameters for all the measurement schemes used in this study.}
\label{tb1}
\begin{tabular}{clll}
\tableline
schemes\tablenotemark{a} & annulus 1 range\tablenotemark{b} & annulus 2 
range\tablenotemark{b}& covered range\tablenotemark{c} \\
\tableline \tableline
3-skip 1$\times$2 lower\tablenotemark{\dag}
		        & $96\fdg5-108\fdg5$  & $193\fdg0-217\fdg0$ &
$18\fdg4-77\fdg6$ \\
4-skip 2$\times$2 lower\tablenotemark{\dag}
		        & $139\fdg8-172\fdg2$ & $139\fdg8-172\fdg2$ &
$0\fdg0-48\fdg8$  \\
4-skip 2$\times$2 upper & $187\fdg8-207\fdg0$ & $187\fdg8-207\fdg0$ &
$0\fdg0-48\fdg8$  \\
4-skip 1$\times$3 lower\tablenotemark{\dag}
		        & $69\fdg3-85\fdg5$   & $207\fdg9-256\fdg5$ &
$46\fdg4-90\fdg0$ \\
4-skip 1$\times$3 upper & $93\fdg5-105\fdg5$  & $280\fdg5-316\fdg5$ & 
$71\fdg2-90\fdg0$ \\
5-skip 2$\times$3 lower\tablenotemark{\dag}
		        & $110\fdg0-134\fdg0$ & $165\fdg0-201\fdg0$ & 
$0\fdg0-60\fdg8$  \\
5-skip 2$\times$3 upper\tablenotemark{\dag}
		        & $150\fdg0-175\fdg2$ & $225\fdg0-262\fdg8$ & 
$0\fdg0-76\fdg0$  \\
6-skip 2$\times$4 lower & $99\fdg5-110\fdg3$  & $119\fdg0-220\fdg6$ &
$16\fdg8-80\fdg0$ \\
6-skip 3$\times$3 lower & $147\fdg0-171\fdg0$ & $147\fdg0-171\fdg0$ &
$0\fdg0-48\fdg8$ \\
6-skip 3$\times$3 upper & $189\fdg0-219\fdg0$ & $189\fdg0-219\fdg0$ &
$0\fdg0-48\fdg8$ \\
8-skip 4$\times$4 lower & $153\fdg0-171\fdg0$ & $153\fdg0-171\fdg0$ &
$0\fdg0-48\fdg8$ \\
8-skip 4$\times$4 upper & $187\fdg8-215\fdg4$ & $187\fdg8-215\fdg4$ &
$0\fdg0-48\fdg8$ \\
8-skip 2$\times$6 lower & $73\fdg5-94\fdg5$   & $220\fdg5-283\fdg5$ &
$42\fdg4-90\fdg0$ \\
8-skip 2$\times$6 upper & $93\fdg5-102\fdg5$  & $280\fdg5-307\fdg5$ &
$62\fdg4-90\fdg0$ \\
\tableline
\end{tabular}

\tablenotetext{a}{The ``lower" or ``upper" in each measurement scheme 
indicates that the lower or upper branch in the time--distance diagram is 
used in this scheme.}
\tablenotetext{b}{Ranges for both annulus 1 and annulus 2 are great-circle
distances from the far-side target.}
\tablenotetext{c}{The distance range from the far-side disk center that can 
be covered by the corresponding measurement scheme.}
\tablenotetext{\dag}{These measurement schemes were previously explored by 
\citet{zha07} and \citet{ilo09}.}

\end{center}
\end{table}

\begin{figure}[!ht]
\epsscale{0.75}
\plotone{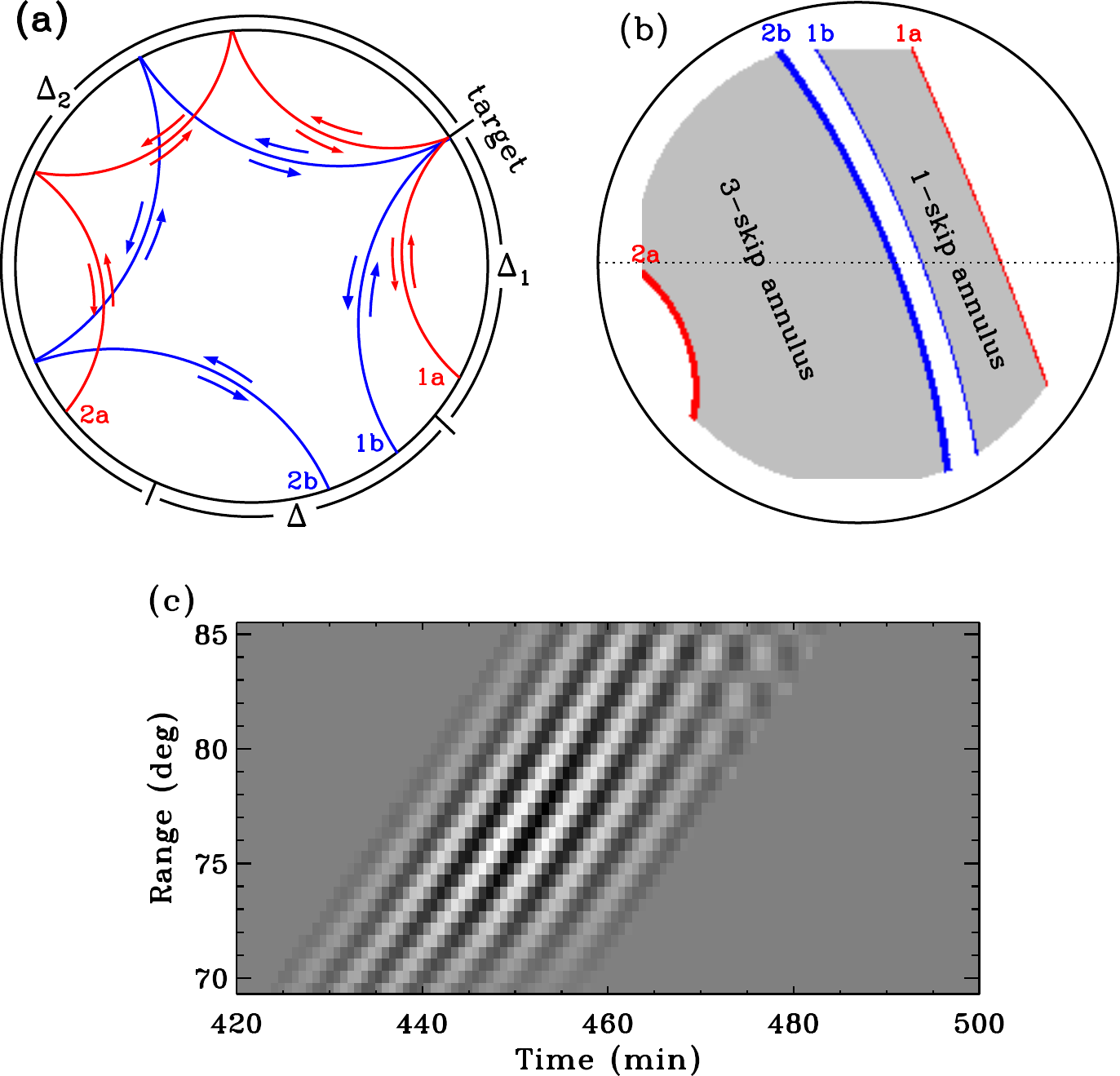}
\caption{ (a) Same as Figure~\ref{f2}c, but with ``1a", ``1b", ``2a",
and ``2b" marked corresponding to the annuli boundaries shown in panel b, and 
with distances $\Delta$, $\Delta_1$, and $\Delta_2$ marked. 
(b) Annuli locations on the near side of the Sun, displayed here in
Postel's projection with a $0\fdg6$ pixel size, for the measurement scheme
shown in panel a. The far-side target is located at longitude of $16\fdg0$ past 
the west limb and latitude of $21\fdg1$N. Red annulus-segment 1a 
(blue annulus-segment 1b) indicates locations with the shortest (largest) 
1-skip distance to the target, and red annulus-segment 2a (blue 
annulus-segment 2b) indicates locations with the shortest (longest) 3-skip 
distance to the target. Thickness of the annulus-segments indicate their 
widths. Shaded regions between 1a and 1b (2a and 2b) indicate the full 
range of 1-skip
annuli (3-skip annuli) used in our calculation. Note that the annuli stop
either at a distance of $60\degr$ away from the disk center or at the boundary
of our datasets. (c) Reference time-distance diagram for the 1$\times$3-scheme, obtained 
after a long-time average and to be used for deriving phase shifts for each 
far-side target. The vertical axis represents 1-skip distance range $\Delta_1$.} 
\label{f2p5}
\end{figure}

The measurement process is rather complicated, and we take the 1$\times$3-scheme 
for the lower branch of 4-skip waves (Figure~\ref{f2}c and Figure~\ref{f2p5}a) 
as one example to 
describe our measurement procedure. The single-skip annulus is taken as 
annulus 1 in Table~\ref{tb1}, the triple-skip annulus is taken as annulus 2, 
and their corresponding measurement parameters are listed in Table~\ref{tb1} 
as well. As shown in Figure~\ref{f2}c, each annulus is delimited by a red 
curve and a blue curve, with the red curve representing the ray path for the 
acoustic waves with the shortest travel distance used in this measurement 
scheme, and the blue curve representing the waves with the longest travel 
distance. In most cases, although the ``annulus" is not a complete annulus 
on the near side but a 
segment of the annulus, we still use the word ``annulus" for simplicity of 
the description. This can be better seen in Figure~\ref{f2p5}b, which is 
plotted corresponding to Figure~\ref{f2p5}a but shows the locations of the 
annuli on the near-side solar disk from the observer's point of view. The solar disk, plotted with a spatial 
resolution of $0\fdg6$ pixel$^{-1}$ in the Postel-projected coordinates, is
consistent with how the data are mapped and selected in our measurements. 
The locations inside annulus-segment ``1a" represent pixels 
with the shortest 1-skip distance of $69\fdg3 - 69\fdg9$ (with the annulus 
thickness of 1 pixel; and also refer to Table~\ref{tb1}) away from the 
far-side target, which is at longitude of $16\fdg0$ past the west limb and 
latitude of $21\fdg1$N, and the locations inside annulus-segment ``2a" are 
pixels with the shortest 3-skip distances of $206\fdg1 - 207\fdg9$ (with 
annulus thickness of 3 pixels) away from the far-side target. Annulus-segments 
``1a" and ``2a" form a pair, inside which the Doppler-velocity signals are averaged 
respectively and then cross-correlated. Similarly, signals are averaged and 
cross-correlated for signals inside the pair ``1b" and ``2b", which show 
locations of signals with the longest travel distance, as well as for all 
other pairs whose distances fall between these two shortest and longest 
pairs within the shaded
areas of ``1-skip annulus" and ``3-skip annulus". For each pair, the 
cross-correlation functions corresponding to both the positive time lags 
and negative lags are averaged to enhance the signal-to-noise ratio. In 
principle, the averaged cross-correlation functions represent, roughly, the 
waveform of waves traveling from one arc on the near side, through the Sun's 
interior and far side, to the other arc on the near side. For one far-side 
target, multiple averaged cross-correlation functions can be obtained 
corresponding to the multiple wave-traveling distances, e.g., results from the 
pair 1a-2a and pair 1b-2b (Figure~\ref{f2p5}b) correspond to a same far-side 
target. 

What is described above is only the measurement procedure for one single 
far-side target. Every different far-side target has different corresponding 
near-side annuli locations, and the same measurement procedure is repeated 
for each far-side target. 
For other measurement schemes shown in Figure~\ref{f2}a and \ref{f2}d, a 
similar procedure is followed; however, the measurement procedure is slightly 
different for the measurement scheme shown in Figure~\ref{f2}b, in which the 
near-side locations usually form a complete annulus. For each travel distance
in this scheme, we divide the whole annulus into four quadrant-annuli, inside 
each of which helioseismic signals are averaged and are cross-correlated with 
the averaged signals from the quadrant-annulus that is directly across.

The measurement schemes shown in Figure~\ref{f2} and the measurement procedures 
described above are essentially the same as the method prescribed by 
\citet{zha07} and \citet{ilo09}. Based on the existing code and the
pre-determined near-side annuli locations corresponding to each far-side 
target location, we further expand the calculation to include more measurement
schemes without significant changes to the existing code and the measurement 
strategy. As already pointed out in Section~\ref{sec21}, measurement schemes 
for 6- and 8-skip waves, namely, 2$\times$4-, 3$\times$3-, 2$\times$6-, and 
4$\times$4-schemes, are therefore adopted in our new code. Moreover,
all waves that experience more than 3 skips have a lower branch and an upper 
branch, but double branches were only explored previously in the 5-skip 
waves; in this study, both branches are included in measurements for all the 
skips. Therefore, a total of 14 measurement schemes, as listed in Table~\ref{tb1}, 
are utilized with only minor code-implementation efforts, resulting in
a great expansion over the previous 5 schemes. 

\subsection{Deriving Phase Shifts}
\label{sec23}
For each far-side target, with a location index $\mathrm{(i,j)}$, and for each 
measurement scheme $s$, a series of cross-correlation functions 
$\mathcal{C}_s^\mathrm{i,j} (\tau, \Delta)$ are obtained as functions of 
travel-time lag $\tau$ and near-surface measurement distance $\Delta$. As shown
in Figure~\ref{f2p5}a,
$\Delta$ is related to $\Delta_1$ and $\Delta_2$, the distances between 
the far-side target and the two separate annulus pairs, by $\Delta = 
360\degr -  (\Delta_1 + \Delta_2)$ for the lower-branch schemes and 
$\Delta = (\Delta_1 + \Delta_2) - 360\degr$ for the upper-branch schemes. 
The $\Delta_1$ and $\Delta_2$ are not independent of each other either, but
are related by $\frac{\Delta_1}{n_1} = \frac{\Delta_2}{n_2}$, where $n_1$
and $n_2$ are the number of skips corresponding to the two annuli.  

The functions $\mathcal{C}_s^\mathrm{i,j} (\tau, \Delta)$ look much like 
their respective lower or upper branch of the corresponding $n$th-skip ridge 
shown in Figure~\ref{f1} or the example shown in Figure~\ref{f2p5}c. To 
derive the travel-time shift (or phase shift) caused primarily by the 
surface-focused far-side target, we take an approach different from the 
prior method \citep{zha07}. In the prior method, the cross-correlation 
function is shifted for each distance by an amount of time, determined from 
either a theoretical expectation or an empirical value measured from a 
long-time average, so that the cross-correlation functions at different 
distances are essentially in phase. Simply put, the wave patterns seen 
in Figure~\ref{f2p5}c are so shifted that they form a straight vertical 
pattern of black and white stripes. One final cross-correlation function, 
which results from averaging all the functions after the time shifting, 
is fitted using a Gabor wavelet function for the travel times. Now we adopt 
a different approach to measure the travel-time shifts, which is believed 
more accurate and more computationally efficient.

All cross-correlation functions $\mathcal{C}_s^\mathrm{i,j} (\tau, \Delta)$ for 
all far-side target locations for each measurement scheme covering a selected 
3-month period (see Section~\ref{sec32}) are obtained following the procedures 
described in Section~\ref{sec22}. For each measurement scheme, we average all 
the available cross-correlation functions. These mean cross-correlation 
functions, which are less noisy and are essentially not shifted by surface 
magnetic field, are used as reference functions $\mathcal{C}_s^\mathrm{ref} 
(\tau, \Delta)$ for the following calculations. Figure~\ref{f2p5}c shows the 
reference cross-correlation function for the 4-skip 1$\times$3-scheme 
lower-branch case. To calculate the phase shift for a far-side target, the 
cross-correlation function for this target at each distance is cross-correlated 
with the reference cross-correlation function for the same measurement distance, 
and this calculation is done in the Fourier domain with frequency dependence 
following the equation:
\begin{equation}
\delta\phi_s^\mathrm{i,j}(\nu, \Delta) = \arg \bigg[ \mathcal{F}\big( 
\mathcal{C}_s^\mathrm{i,j} (\tau, \Delta) \big) \cdot \mathcal{F}^\dagger\big( 
\mathcal{C}_s^\mathrm{ref} (\tau, \Delta) \big) \bigg],
\label{eq1}
\end{equation}
where symbol $\mathcal{F}$ represents Fourier transform, $\nu$ is frequency 
after the Fourier transform is applied on time $\tau$, $^\dagger$ represents 
complex conjugate, and $\arg$ takes the argument of the complex number. That is, 
the travel-time shift (phase shift) carried by a cross-correlation function is 
assessed through comparing with a reference function, a method recently 
explored and utilized by \citet{che18} for a different study. In the Fourier 
domain, the relative phase-shift, which is the argument of the complex number, 
can be obtained for frequencies between 3.0 to 5.0 mHz. The phase shifts for 
multiple frequencies are averaged again for the phase shift corresponding to this 
far-side location and for this measurement distance. The same procedure is 
repeated for different measurement distances, and the mean phase-shift from 
all distances for this measurement scheme is then used as the relative 
phase-shift for this far-side target location, i.e.,
\begin{equation}
\delta\phi_s^\mathrm{i,j} = \langle \langle \delta\phi_s^\mathrm{i,j} (\nu, 
\Delta) \rangle_\nu \rangle_\Delta,
\label{eq2}
\end{equation}
where $\langle \ \rangle_\nu$ takes an averaging over all $\nu$'s, and $\langle
\ \rangle_\Delta$ averages over all $\Delta$'s. The above procedure is then 
repeated for all far-side locations $\mathrm{(i,j)}$ and repeated for all 
measurement schemes $s$. In the end, each measurement scheme gives one far-side 
map of acoustic phase shifts, although none of the measurement schemes (except 
the 5-skip 2$\times$3-scheme) gives a map covering the entire far-side Sun.

\section{Data Analysis and Results}
\label{sec3}

\subsection{Data Preparation and Processing}
\label{sec31}

Only low- and medium-$\ell$ helioseismic modes are needed for imaging the 
Sun's far-side ARs, therefore, the high-resolution Dopplergrams observed 
by \sdo/HMI need to be pre-processed to reduce their spatial resolution before 
being used in our far-side imaging calculations. In our new code, we use 
the HMI vector-weighted data \citep[data series ``hmi.vw\_V\_45s" in the 
\sdo/JSOC webpage;][]{lar18}, which mimic {\it SOHO}/MDI's structure program 
data \citep{kos97, lar13} and have a pixel size of about $0\fdg60$. The 
full-disk data of this type are tracked with the Carrington rotation rate 
and a temporal cadence of 45 sec, the same as the observational cadence.
For the tracking, we select 00:00UT and 12:00UT of each day as the middle 
time of a tracked period of 1860 min (31.0 hr), which is composed of 
2480 \sdo/HMI Dopplergrams. Each tracked full-disk Dopplergram is then 
converted to a Postel-projected map, with the apparent disk center at the 
middle time of the tracked period as the projected map center, so that 
the annual variation of the solar $B$-angle can be neglected in our 
calculation. The remapped image keeps a spatial resolution of 
$0\fdg60$~pixel$^{-1}$. To examine the validity and robustness of using the 
vector-weighted low-resolution data as our data input, we have tested 
tracking, rebinning, and remapping HMI's original data, and found that 
the full-resolution data give us essentially the same final far-side maps, 
but substantially slow down the processing.

As pointed out by \citet{zha07}, to analyze the very weak signals of the
waves that start from the near side of the Sun and end in the near side after 
traveling to the far side, filtering out high-$\ell$ modes is necessary to
enhance the signal-to-noise ratio of the far-side maps. Generally, the 
acoustic modes covering a frequency range of $2.5 - 5.0$ mHz and an $\ell$ 
span of $3 - 50$ are used, although the detailed filtering parameters 
slightly change from one measurement scheme to another, depending on the 
number of wave skips as well. This filtering process is done after Fourier 
transform is applied on our tracked datacube. Strictly speaking, using 
Fourier transform for filtering is not robust because distances between most 
locations are not precisely preserved in the Postel-projected maps; however, 
our test analysis using the spherical-harmonics-decomposed data showed that 
Fourier transform in this study is an acceptable compromise between quality 
and computation speed. 

\begin{figure}[!ht]
\epsscale{0.65}
\plotone{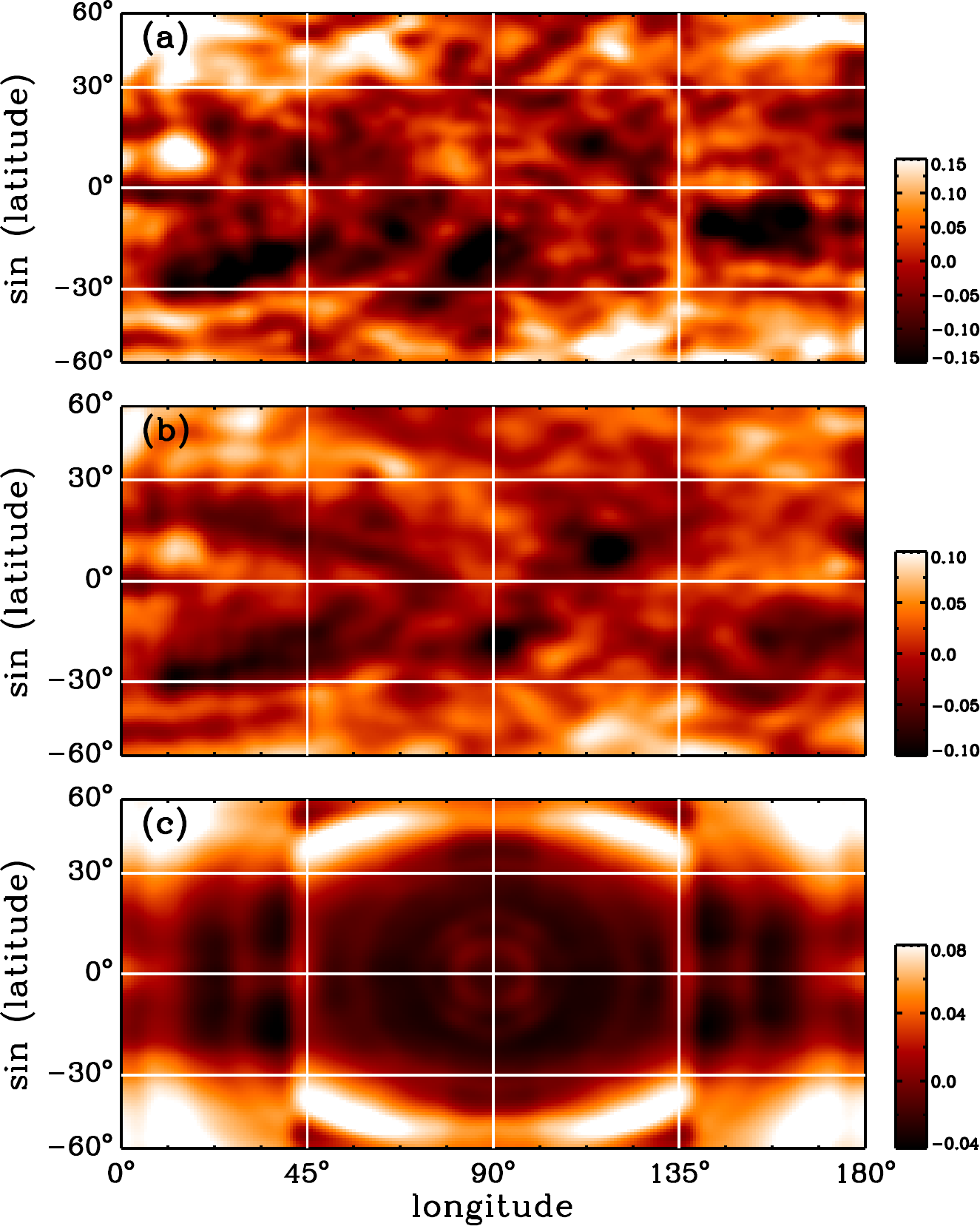}
\caption{(a) Time--distance far-side image, obtained from a 31-hr period with
a middle time of 00:00UT 2014 March 13, after averaging all available images 
that result from the 14 measurement schemes. (b) Far-side image after the 
image in panel (a) is overlapped and averaged with images of 3 prior days. 
(c) Background far-side image, averaged from all far-side images obtained 
during the analyzed 3-month period, but before the images are overlapped 
with images of prior days. }
\label{f3}
\end{figure}

\subsection{Far-Side Maps of Active Regions}
\label{sec32}

Following the analysis procedure described in Section~\ref{sec2}, we have 
processed 9-year-worth of \sdo/HMI observations for the period of 2010 May 1
through 2019 July 31. In this study, we only choose the results of a 
3-month period, covering 2014 January 1 to March 31, to examine the far-side 
imaging quality of our new code. For each of the tracked datacubes, a total 
of 14 far-side images are obtained, all of which (except the 5-skip 
2$\times$3-scheme that gives a full coverage) cover either the central 
part or both limb areas of the far-side Sun.  Examining all 
images that are produced for the whole 3-month period, we find that it is 
unlikely that all the 14 far-side images can image the far-side ARs equally 
well for any given period. Some measurement schemes give systematically 
better imaging quality than other measurement schemes, but no specific 
measurement scheme can always map the far-side ARs better than all other 
schemes. However, it is true that our final far-side images, made from 
merging all the 14 measured images, possess consistently better quality than 
any images made from one single measurement scheme, and are nearly always 
better than the images made from our previous codes \citep{zha07, ilo09}.

The final far-side image is an average of all the 14 far-side images, with 12 
of them having an equal weight and the other 2, i.e., the schemes of 
``3-skip 1$\times$2 lower" and ``8-skip 2$\times$6 upper", assigned half 
weights of other schemes. These two schemes consistently give lower-quality 
images, and are hence assigned lower weights. Different areas of the far 
side are covered by a varying number of the 14 individual images, with 
a minimum coverage of 6 times close to the far-side limbs and a maximum of 
14 times in the areas that join the central and limb regions around $50\degr$ 
from the far-side disk center. Other far-side areas receive a coverage 
of 8 times. Figure~\ref{f3}a shows an example of far-side images made over 
a 31-hr period centered at 00:00 UT, 2014 March 13. Negative phase shifts 
(dark patches in the plot) are caused by the magnetic field of the far side, 
hence representing far-side ARs.

\begin{figure}[!h]
\epsscale{0.75}
\plotone{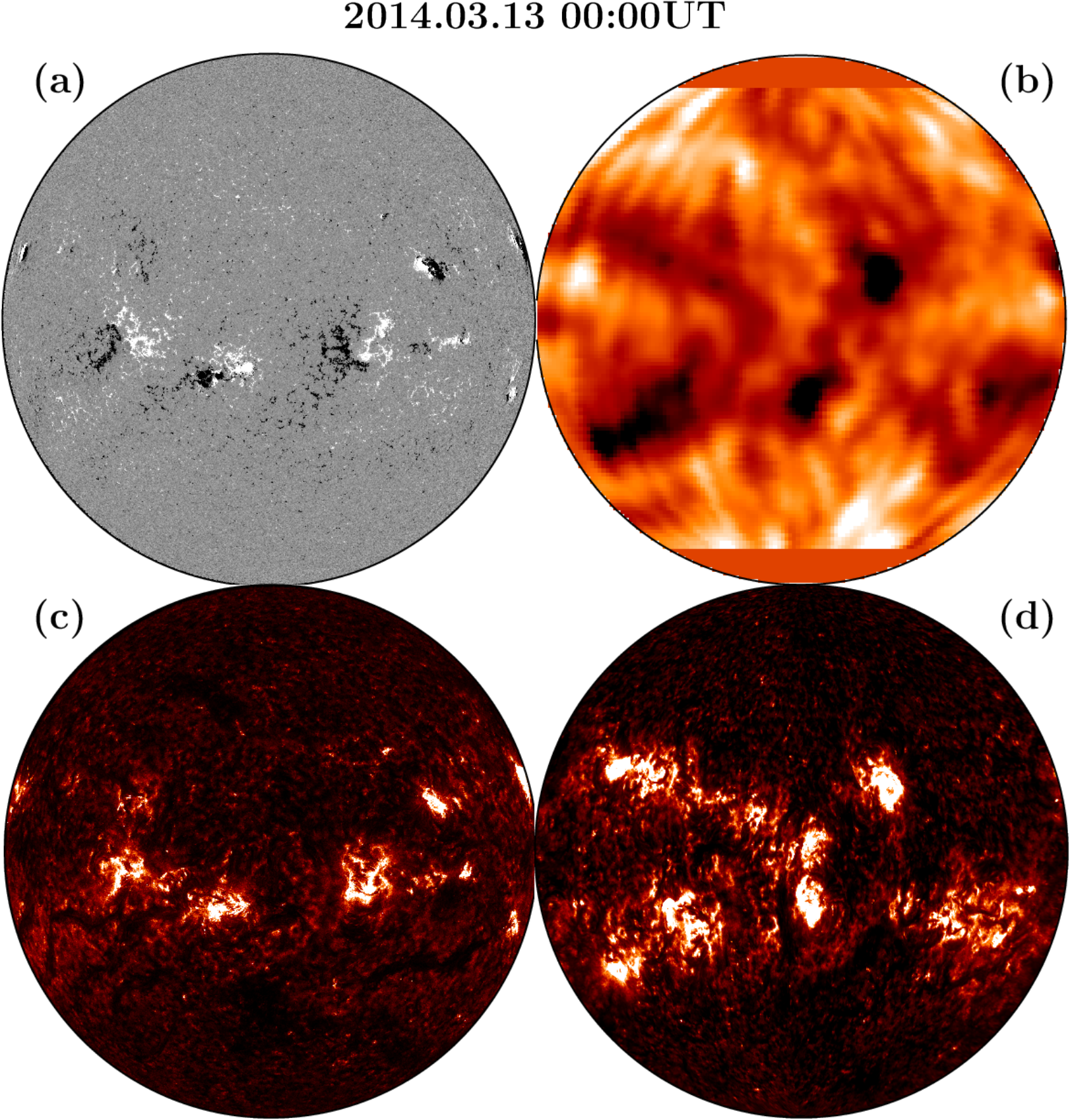}
\caption{Comparison of near-side and far-side images taken at or near 00:00 
UT of 2014 March 13. (a) \sdo/HMI-observed near-side magnetic field; (b) 
helioseismic far-side image, same as Figure~\ref{f3}b but displayed in 
a different coordinate frame; (c) \sdo/AIA-observed near-side 304~\AA\ image; 
(d) {\stereo}/EUVI-observed far-side 304~\AA\ image after merging observations 
from both {\stereo}-A and {\stereo}-B spacecraft. The magnetic field is 
displayed with a scale of $-$200 to 200 Gauss, the helioseismic image is 
displayed with a scale of phase shifts from $-$0.10 to 0.10, and the EUV 
images are displayed with a DSN from 10 to 2000. The \stereo/EUVI 304~\AA\ image 
is displayed after its intensity is calibrated to match the \sdo/AIA 304~\AA\ 
level.}
\label{f4}
\end{figure}

Despite the great improvement our new helioseismic far-side imaging code has 
made over the previous codes, it is also acknowledged that far-side images 
made from the 31-hr periods can only map the shape and size of far-side ARs 
very roughly.  The shape and size of an AR can change significantly from one 
imaging period to the next, beyond the reasonable temporal evolution one 
would expect. To enhance the stability and reliability of their far-side images, 
Lindsey et al.\footnote{\url{http://jsoc.stanford.edu/data/farside/}} 
proposed to overlap one far-side image with the far-side images taken from 
4 prior days (a total of 120 hours, or 5 days, of integration time), a process 
that sacrifices temporal resolution for image stability. In this study, we 
follow a similar approach but overlap one far-side image with the images 
obtained from 3 prior days (a total of 103 hours integration time considering 
each image is made from a 31-hr period), i.e., an average of 7 individual 
images after a uniform solar rotation (i.e., not differential rotation) is 
accounted for. However, not every far-side location has prior images to 
average with: the majority of far-side areas are averaged over 7 images, 
while the areas closer to the west limb are averaged fewer times. 
Figure~\ref{f3}b shows an example far-side image, obtained after 
the image in Figure~\ref{f3}a is averaged with images of 3 prior days; 
Figure~\ref{f4}b shows the same image in a different coordinate system. The 
overall shape, size, and strength (i.e., the amount of helioseismic-wave 
phase-shifts) of most ARs get altered after such averaging, but these ARs and 
their physical parameters become more consistent and stable from day to day, 
allowing us to potentially calibrate the far-side helioseismic maps into 
magnetic flux maps in the future.

Figure~\ref{f3}b and \ref{f4}b are displayed with a scale of $-0.10$ to $0.10$ 
radians in the measured phase shifts, which correspond to a travel-time 
shifts of about $-5.3$ to $5.3$ sec. This display scale is selected arbitrarily, 
which unavoidably impacts how the far-side images of ARs are perceived in their
size and strength. However, the selection of this display scale allows a 
better match, albeit only visually, of the helioseismic far-side ARs with 
the \stereo/EUVI 304~\AA\ images of ARs (see Section~\ref{sec4}).

We also acknowledge that despite our new efforts, artifacts still remain in 
the new far-side maps of ARs. Figure~\ref{f3}c shows a background image, 
obtained after averaging all the far-side images of the three analyzed months 
of 2014 before the images are shifted and overlapped for a more stable image 
quality. As already mentioned, phase shifts obtained at different far-side 
locations result from different measurement schemes using acoustic signals 
of slightly different wave modes observed at different near-side locations; 
therefore, it is not surprising that the phase shifts vary with far-side disk 
location in addition to the AR-caused shifts. Generally, the phase shifts 
tend to be smaller in low-latitude areas and larger in the near-limb and 
higher-latitude areas, where white (positive) patches appear. This is likely 
due to the near-side systematic center-to-limb effect in helioseismic 
measurements \citep{zha12} leaking into the far-side images. It is not 
straightforward to remove this effect, at least not by simply subtracting 
this mean background image from every far-side image, because this would 
cause artificial ARs near the limb and high-latitude areas in many far-side 
images. Fortunately, only dark patches corresponding to negative phase shifts 
are useful in identifying far-side ARs, and those white patches can be 
neglected for the time being.

\section{Comparisons}
\label{sec4}

\subsection{Comparisons with \stereo\ Far-Side EUV Images} 
\label{sec41}

As introduced in Section~\ref{sec1}, two \stereo\ spacecraft together 
monitored the Sun's entire far-side activity during February 2011 and October 
2014. Particularly, the 304~\AA\ images observed by \stereo/EUVI are 
highly correlated with maps of magnetic flux \citep{uga15}, giving us valuable 
pictures of the far-side ARs that can be used to assess the reliability 
of helioseismic far-side imaging techniques \citep{lie17}.  Figure~\ref{f4} 
shows a comparison of the Sun's \sdo/HMI-observed near-side magnetic field 
and \sdo/AIA-observed EUV 304~\AA\ flux, together with a far-side helioseismic 
image from our new method and \stereo/EUVI-observed 304~\AA\ image, all taken 
at about a similar time 00:00UT of 2014 March 13 (keep in mind that the 
helioseismic image results from a 103-hr average). 

The comparison of Figure~\ref{f4}a and \ref{f4}c shows the good correlation 
between the ARs' magnetic flux and their 304~\AA\ brightness enhancements,
implying that the areas of 304~\AA\ brightening can be used as a good indicator 
to the locations and sizes of ARs. The comparison between Figure~\ref{f4}b 
and \ref{f4}d is a representation of how the far-side helioseismic images 
are comparable with the 304~\AA\ images of ARs. It can be seen that the 
helioseismic method is capable of mapping most of the major ARs, with a 
good location and size indication, but loses detailed structures of these 
regions due to the poor spatial resolution resulting from the long helioseismic 
wavelengths used in our method. It is likely that the phase shifts measured 
in these far-side ARs and the EUV 304~\AA\ flux density observed for the 
same ARs have a good correlation, but this needs further confirmation. 
Overall, Figure~\ref{f4} demonstrates a promising potential that the far-side 
helioseismic images can be calibrated into far-side magnetic-flux maps 
by using the \stereo's 304~\AA\ images as a stepping-stone.

\begin{figure}[!ht]
\epsscale{0.8}
\plotone{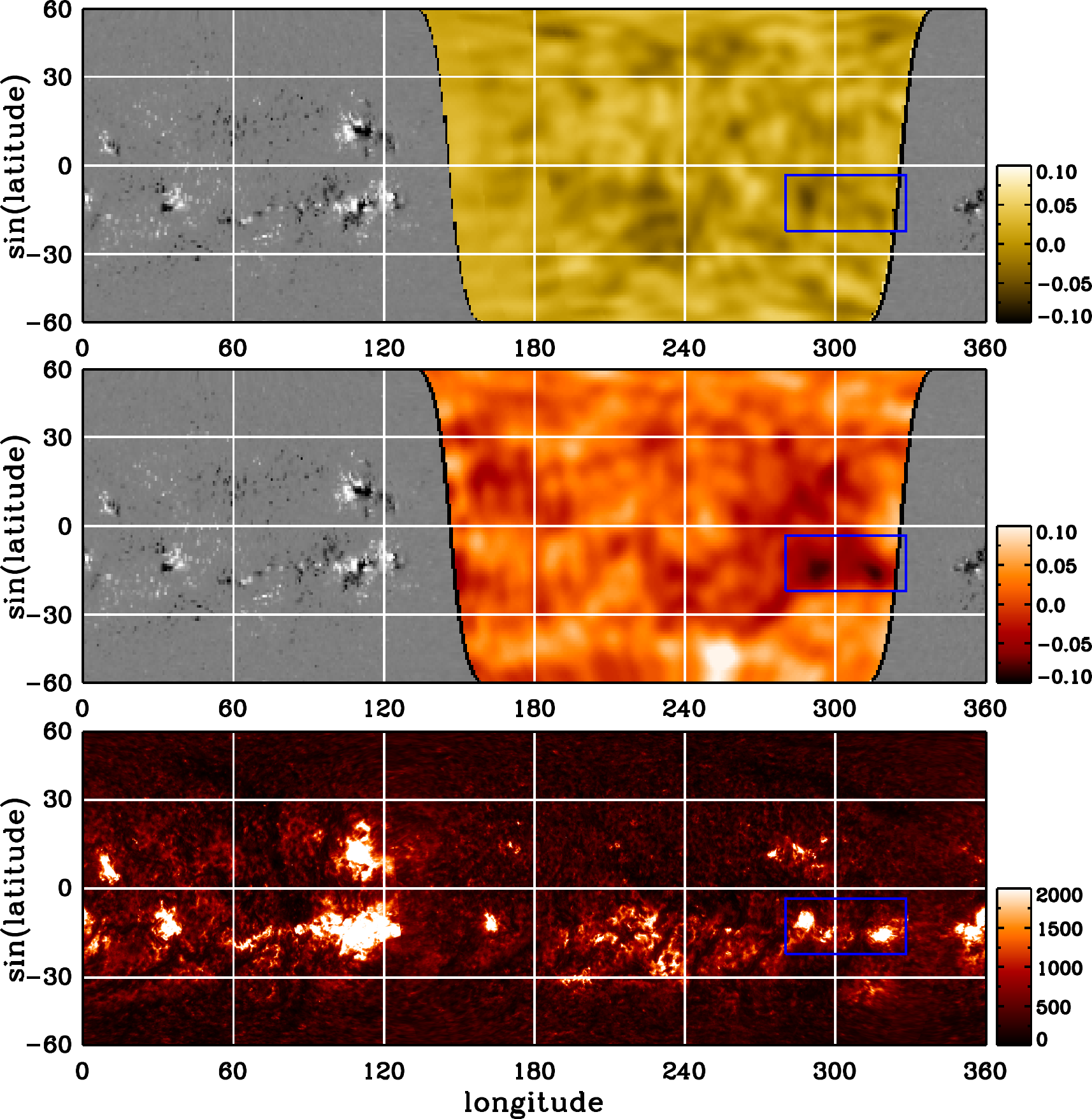}
\caption{Comparison of the helioseismic holography far-side image, our new 
time--distance far-side image, and the EUV images taken at 12:00UT, 2014 
February 7. Top panel shows a synchronic chart of \sdo/JSOC helioseismic 
holography far-side image ({\it color}) and \sdo/HMI near-side magnetic field 
({\it black and white}); middle panel is a similar chart composed of 
time--distance far-side image ({\it color}) and near-side magnetic field; and 
bottom panel shows a combination of simultaneous 304~\AA\ observations of 
\sdo/AIA and both \stereo\ spacecraft. All images are mapped into the same 
Carrington longitude -- sin(latitude) coordinates, and only latitudes between 
$-60\degr$ and $60\degr$ are displayed. Magnetic field is displayed with a 
range of $-$200 to 200 Gauss. Areas delimited in the blue boxes are chosen 
for comparisons.}
\label{f5}
\end{figure}

\begin{figure}[!ht]
\epsscale{0.8}
\plotone{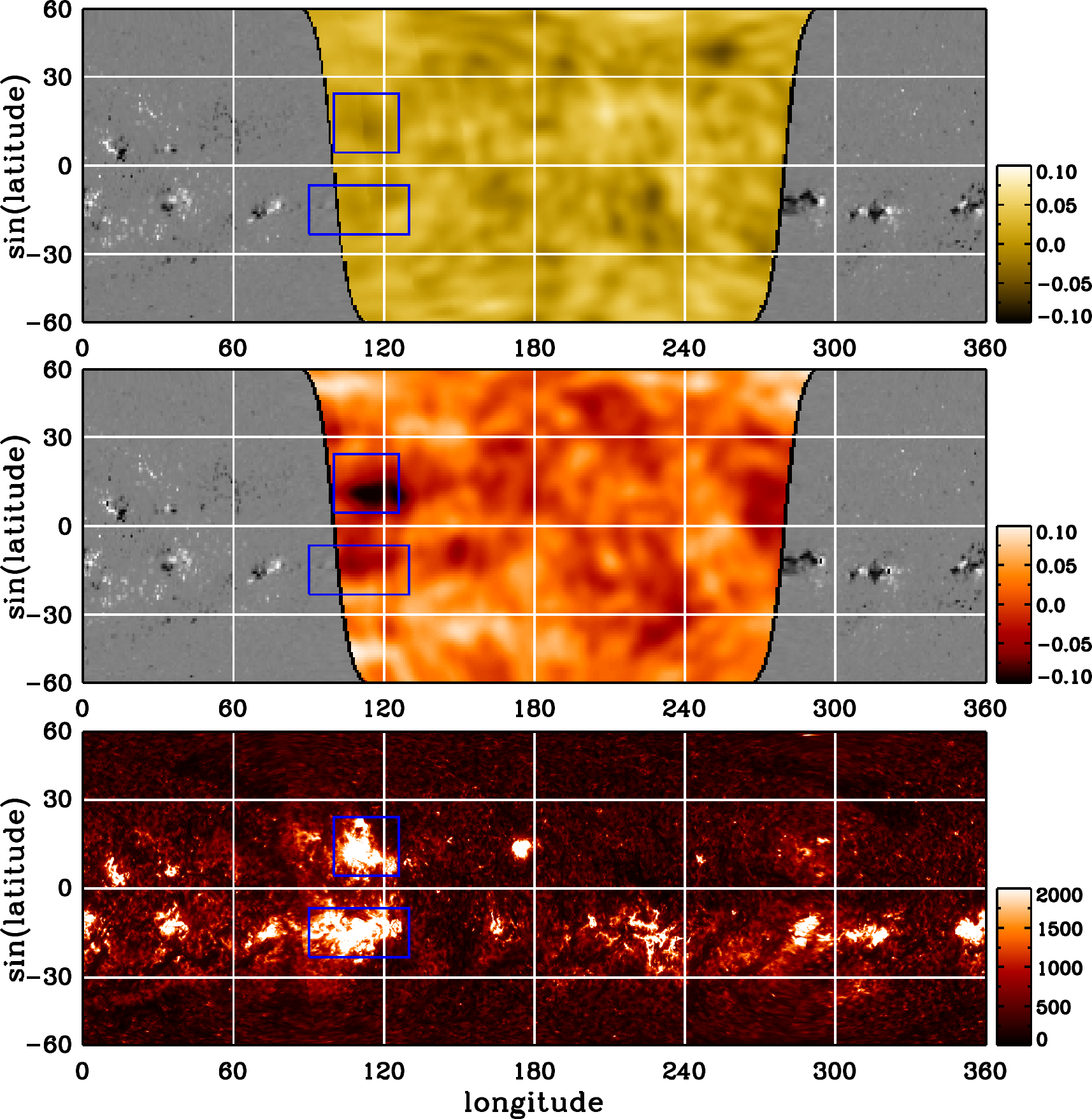}
\caption{Same as Figure~\ref{f5}, but for 00:00UT, 2014 February 11.}
\label{f6}
\end{figure}

\begin{figure}[!ht]
\epsscale{0.8}
\plotone{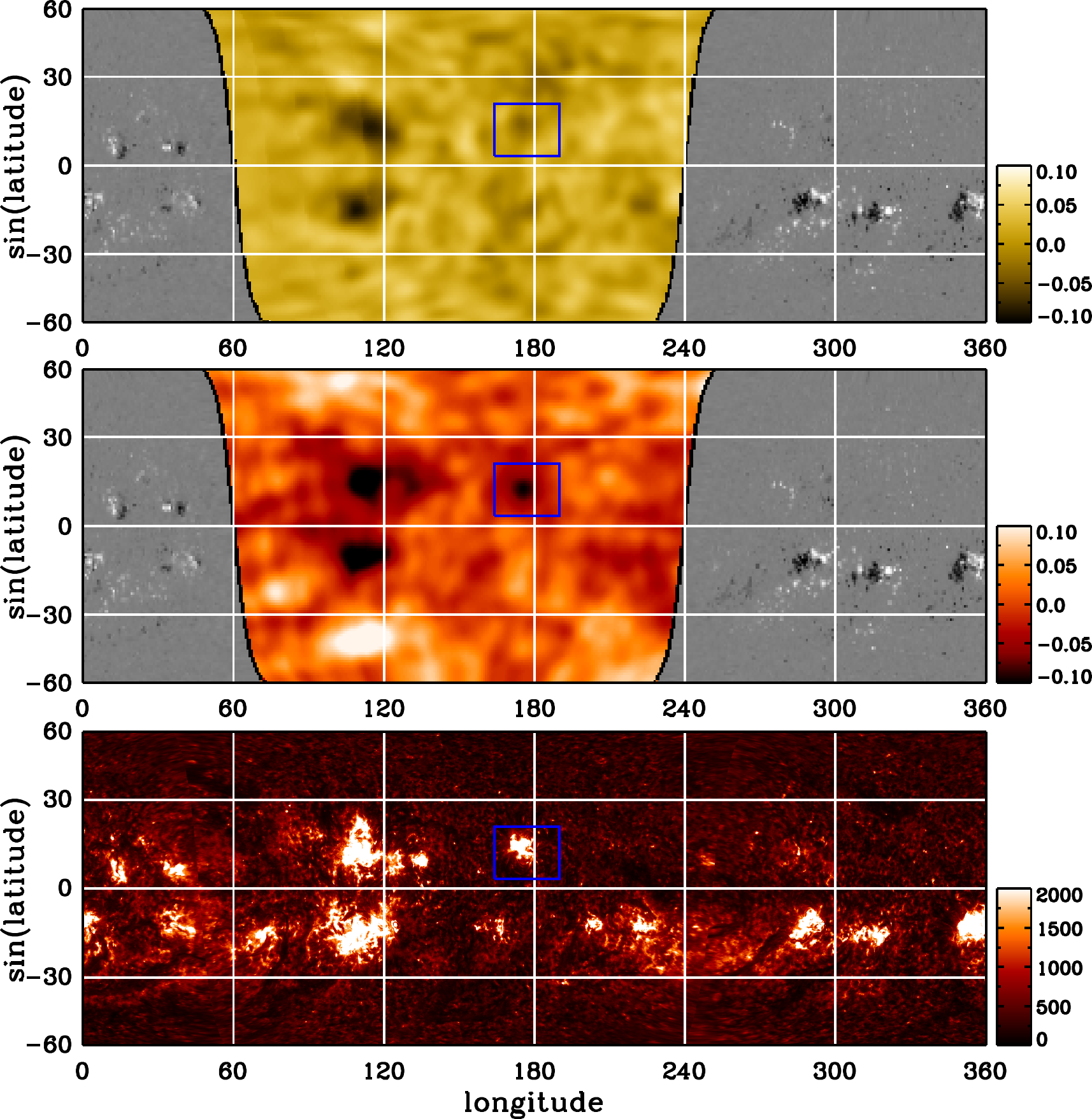}
\caption{Same as Figure~\ref{f5}, but for 00:00UT, 2014 February 14.}
\label{f7}
\end{figure}

\subsection{Comparisons with Helioseismic Holography Far-Side Images}
\label{sec42}

Including more time--distance helioseismic measurement schemes unsurprisingly 
gives us better far-side imaging quality than the previous time--distance 
far-side imaging method \citep{zha07, ilo09}. Meanwhile, it is also desirable 
to compare our new results with the routine \sdo/JSOC far-side images 
produced by the helioseismic holography method, which has also been upgraded 
in recent years to better accommodate the \sdo/HMI data as well as to improve 
the imaging quality. Figures~\ref{f5} -- \ref{f7} show a few representative 
comparisons of the images from the two helioseismic methods and the 
corresponding \sdo/AIA and \stereo/EUVI images, highlighting a few selected 
ARs: near the east limb not long before they rotate onto the near side, 
near the west limb not long after they rotate onto the far side, and 
not long after a far-side AR emergence. In these figures, the data for both 
the helioseismic holography images and magnetic fields are downloaded from 
\sdo/JSOC website, and the data for the composite 304~\AA\ images were prepared 
by \citet{lie17} and kindly provided to us by the authors.

The capability of predicting arrivals of large ARs, which rotate into the
Earth's view from the Sun's east limb, is practically useful for space weather 
forecasting. Figure~\ref{f5} highlights two ARs, delimited in the blue box,
about 1 day before they appeared on the near side. These two ARs would become 
NOAA AR11976 and AR11977 after their near-side appearance.  Both the 
helioseismic holography and time--distance images are displayed with a same 
scale of $-$0.10 to 0.10 phase-shift, which is consistent with the display
scale of the online holography images at the \sdo/JSOC website. It can be 
seen that the time--distance image unambiguously detects the two highlighted ARs, 
which approximately agree in location with the brightening areas (presumably
ARs) observed in 304~\AA. The location difference between the time--distance 
helioseismic images and the 304~\AA\ observations falls within the spatial 
uncertainties of the helioseismic images, which can be as poor as $10\degr$. 
That the helioseismic holography fails to show these ARs is not due to the 
display scale of the image, because the signals inside the blue box do not 
exceed the variation level of the background.

Figure~\ref{f6} shows an example of two other ARs, AR11967 in the southern 
hemisphere and AR11968 in the northern hemisphere, about 1 day after their rotation 
from the Sun's west limb onto the far side. As pointed out in Section~\ref{sec3}, 
both the holography and time--distance methods show the far-side 
near-west-limb areas with fewer days' image overlapping, unlike most other 
areas on the far side. It is therefore expected that ARs in this area 
are not as reliably mapped. Nevertheless, the time--distance method shows 
one of the two ARs (the upper one) unambiguously, whereas the holography 
method does not show either AR convincingly.

Other than the ARs near both the far-side limbs, both helioseismic holography
and time--distance methods give essentially consistent results in most areas 
of the far side, except for emerging ARs. Undoubtedly, new ARs also emerge on 
the far side, and some of them emerge rapidly and become strong and large enough 
to be detectable through helioseismic methods not long after their emergence. 
Figure~\ref{f7} shows such an example. According to the \stereo\ observations 
that do not require a days-long integration like the helioseismic methods, 
the magnetic flux emergence started before February 11, when it is already 
visible in the 304~\AA\ map in Figure~\ref{f6}, but only became unambiguously 
visible in the time--distance far-side images 3 days later on February 14. 
The holography method picks up the signal of this region unambiguously 
two more days later on February 16. The reason that the time--distance method only 
detects this emerging region over 3 days after its emergence is two-fold: 
first, as will be discussed in more detail in Section~\ref{sec5}, helioseismic 
methods are only sensitive to the ARs that are larger than a certain surface 
area and stronger than a certain magnetic-field strength; second, averaging 
with 3-day prior images smears temporal resolution of the helioseismic images. 
It is possible that our method detects this emerging region 1 or 2 days 
earlier, but the region becomes ambiguous after averaging with earlier images 
that do not detect this AR.

These above examples demonstrate that our new far-side imaging code has an 
advantage over the helioseismic holography method in the detection of newly 
emerged ARs and ARs near both the far-side limbs.

\begin{figure}[!t]
\epsscale{1.0}
\plotone{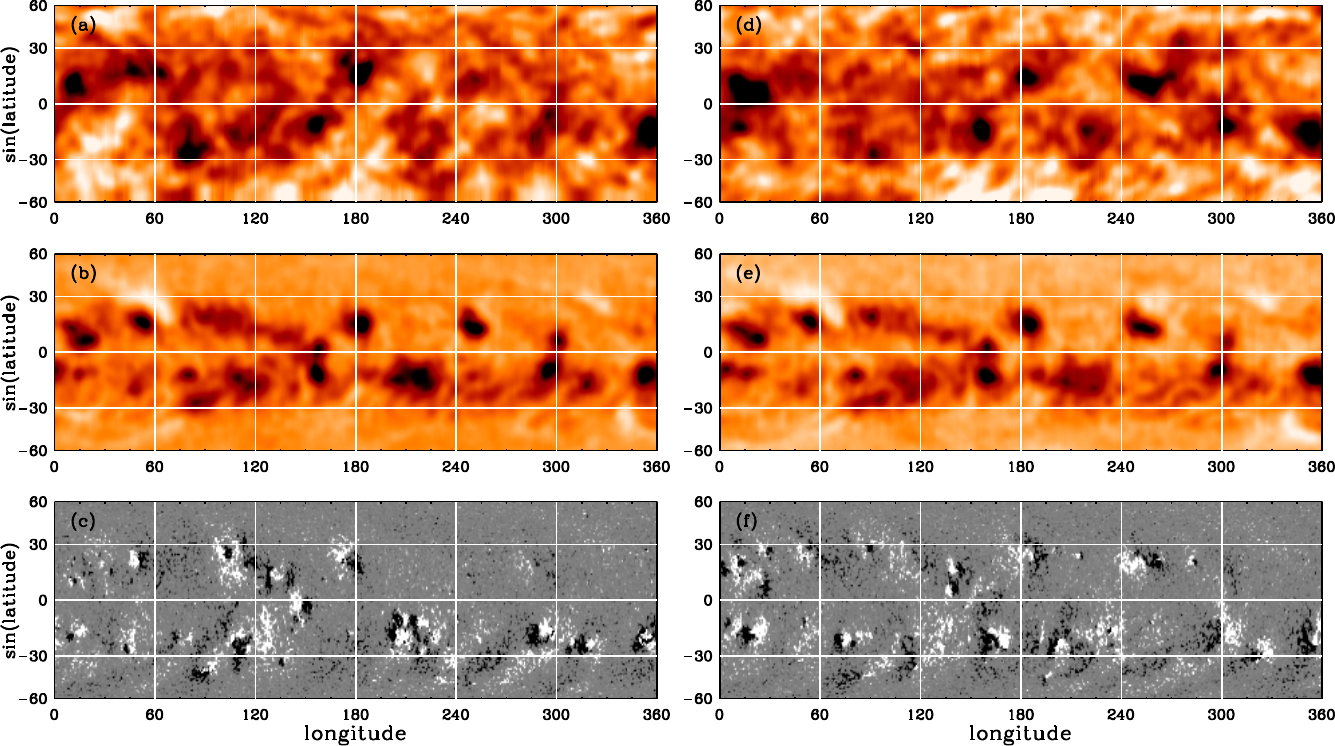}
\caption{Synoptic maps of (a) time--distance helioseismic phase shifts using 
data near the far-side central meridian, (b) \stereo/EUVI 304~\AA\ flux using 
the data near the far-side central meridian, displayed in the negative of 
the logarithm of the flux, (c) near-side magnetic field of Carrington rotation 
(CR) 2147.  (d-e) Same as panels (a-b), but made using the data $60\degr$ east of 
the far-side central meridian. (f) Same as panel (c) but for CR2148. Panels (a) 
and (b) use the data of 2014 February 26 through March 23, and panels (d) and (e) 
use the data of March 2 through 27. CR2147 covers the period between February 11 
and March 10, and CR2148 covers March 10 through April 6. The far-side 
helioseismic maps are displayed with a range of $-0.08$ to 0.06, the far-side
EUV images are displayed with a range of $-7.5$ to $-5.0$, and magnetic field
charts are displayed with a range of $-100$ to 100 Gauss.}
\label{f8}
\end{figure}

\subsection{Synoptic Far-Side Maps}
\label{sec43}

Rather than selected snapshots with only one or two ARs, which may happen to 
be well or badly mapped, a Carrington synoptic map made from one entire 
rotation of the Sun can give us a more general perception of the reliability 
of the helioseismic far-side images. Figure~\ref{f8} shows a comparison of 
a couple of pairs of synoptic maps made from far-side helioseismic images 
and \stereo\ 304~\AA\ images. One pair of the data, shown in Figure~\ref{f8}a 
and \ref{f8}b, from the helioseismic and the EUV images, respectively, are 
made by overlapping $30\degr$-wide longitudinal bands across the far-side 
central meridian using two images per day; and another pair, shown in 
Figure~\ref{f8}d and \ref{f8}e, are made by overlapping $30\degr$-wide 
bands across the $60\degr$ longitude east of the far-side central meridian. 
Figure~\ref{f8} also includes two near-side synoptic maps of magnetic field, 
obtained from $30\degr$-wide bands across the near-side central meridian 
for the Carrington rotation before the same regions rotate onto the far side 
and after they rotate out of the far side. That is, one set of the far-side 
synoptic maps is made half of a rotation after the same areas are used to 
make the first near-side synoptic magnetic map, and another set just $30\degr$ 
before they rotate back into the near side. It is worth pointing out that 
the central area of the helioseismic far-side images primarily result from 
the 4-skip 2$\times$2, 6-skip 3$\times$3, and 8-skip 4$\times$4 measurements 
schemes, and the limb areas are primarily made using other measurement 
schemes; therefore, these two sets of synoptic maps (Figure~\ref{f8}a and 
\ref{f8}b) essentially show us the sensitivity and reliability of different 
schemes.

Comparing Figure~\ref{f8}a and \ref{f8}b, and Figure~\ref{f8}d and \ref{f8}e,
we find that many ARs from the helioseismic images and the EUV maps agree
with each other, although the size, shape, and exact location differ in most 
cases. However, there are a couple of clear exceptions at coordinates of 
approximately ($210\degr$, $-15\degr$) and ($250\degr$, $10\degr$), where 
the helioseismic ARs are largely missing in Figure~\ref{f8}a. However, if 
we refer again to Figure~\ref{f8}d that is made with different
measurement schemes using the data of about 5 days later, we find that the 
region at ($250\degr$, $10\degr$) is actually a new emerging region that 
was still growing when the data for Figure~\ref{f8}a were taken. The region 
at ($210\degr$, $-10\degr$) is just the opposite, fast decaying when it 
was near the central meridian based on the EUV map. However, despite its 
weakened strength, it becomes detectable in Figure~\ref{f8}d, implying 
that weaker ARs are easier to be detected closer to the east limb than 
in other far-side locations. This is very useful because ARs closer
to the east limb are more likely to cause space-weather effects as they
rotate toward the near side.

We again caution that only negative phase shifts in the helioseismic 
far-side maps are meaningful in identifying and possibly calibrating 
ARs, and the positive patches in such maps are mostly due to the systematic
center-to-limb effect in our local helioseismology technique, hence cannot 
be interpreted as AR-related or other surface- or interior-related features.

\section{Success Rate}
\label{sec5}

In Section 4, through comparing far-side images obtained from our new helioseismic 
code with the \stereo/EUVI observations of 304~\AA\ flux, we see how 
helioseismic images of ARs agree or disagree with the EUV brightening areas of ARs. 
However, a statistical assessment of success rate of our new helioseismic 
detection, just as what \citet{lie14, lie17} have done for the helioseismic 
holography far-side images, is necessary. Here, we use the same 3-month 
period of helioseismic data used in Sections~\ref{sec3} and \ref{sec4}, and 
the contemporaneous composite 304~\AA\ maps as shown in Figures~\ref{f5} -- 
\ref{f7} for such a statistical assessment.

\subsection{Data Preparation}
\label{sec51}

Each of our helioseismic far-side images has only $225\times121$ pixels, 
covering $180\degr$ in longitude and $120\degr$ in latitude (we use sin(latitude) 
in calculation, though), and each snapshot image results from an average of 
103-hr observations. Therefore, for a fair and reasonable assessment of 
helioseismic images against the 304~\AA\ observations, two steps need to 
be taken: first, lower the spatial resolution of the 304~\AA\ images to match 
that of the helioseismic images; and second, average the 304~\AA\ images 
with the images of prior days in the same way the helioseismic images are 
averaged.

The composite 304~\AA\ maps, prepared and provided to us by \citet{lie17}, cover 
the entire Sun with $3600\times1800$ pixels and are in longitude -- latitude 
coordinate. We first take a logarithm of the EUV flux values, so that small 
areas of high flux density do not spread to larger areas in the process of 
reducing the spatial resolution. Then we reduce the spatial resolution by 
convolving each EUV image with a two-dimensional Gaussian function that has 
a FWHM of $4.71\degr$. The selection of the width of the Gaussian function 
is empirical: we tested a number of different FWHM parameters and selected 
the one that best visually matches the helioseismic images. However, we also 
acknowledge that given the apparent differences between the EUV images and 
helioseismic images, as can be seen in Figures~\ref{f4}-\ref{f8}, the ``best 
match" is poorly defined. 

The spatially-reduced EUV images are then remapped from the longitude -- latitude 
coordinates to longitude -- sin(latitude) coordinates to match the helioseismic 
images through bilinear interpolation. The Carrington longitude covers $180\degr$ (more 
or less, depending on the solar $B$-angle) of the far side; only observations 
between $-60\degr$ and $60\degr$ latitude are kept for further analysis, 
because no ARs appear higher than latitudes of $60\degr$. The EUV images are 
then rebinned to the same number of pixels of the helioseismic images, 
each pixel having a size of $0\fdg8\times0\fdg8$, or 0.64 square degrees (the 
degree and square degree correspond to the pixel sizes at the disk center). 
Such pixels, each equivalent to approximately 31.0 $\mu$hemispheres, will 
be referred as helioseismic-image pixels (HIPs) hereafter.

To match the temporal averaging of the helioseismic images, each EUV image 
is overlapped and averaged with images of 3 prior days, using two images of 
12-hr apart each day. Thus, the final resultant EUV images have both spatial 
and temporal resolutions matching those of the helioseismic far-side images, 
and these two types of images can then be overlapped for a comparative 
statistical analysis.

\subsection{Matching Criterion}
\label{sec52}

There are a total of 180 far-side helioseismic images, and the same number of 
EUV images, during the studied three months. Each helioseismic and EUV far-side 
image, despite resulting from 103-hr and 3.5-day average, respectively, is 
considered a random snapshot of the Sun's far side and compared individually. 
That is, unlike the approach of tracking one AR across the far-side disk 
\citep{lie17}, we consider an AR in one snapshot as one individual AR, 
although the same AR almost certainly appear on the following images. This 
way, many successfully-detected ARs are counted multiple times, and meanwhile, 
those undetected ARs may also be counted as failures multiple times. This may seem 
to cause biased results; however, because any individual image can be thought 
as a snapshot from a random day, there are a sufficient number of ARs through 
the 90-day period to make the statistics unbiased. Also, a single AR may 
not be successfully detected throughout its lifetime on the far side, as 
seen in Figure~\ref{f8}, so this approach helps take into account the number 
of sporadic failed detections which would otherwise be considered complete 
failures.

None of the far-side ARs have a NOAA-denominated active region number, thus 
the boundaries of these ARs are not defined. After the reduction of the EUV 
flux maps' spatial resolution, it is possible that two or more adjacent ARs 
cannot be clearly distinguished with a distinct separation. In such a case, 
we count the ARs as one in this study. After convolving with the Gaussian 
function, the areas of EUV ARs become substantially larger and boundaries 
become more blurred. Therefore, it is highly likely one AR in our counting 
is composed of more than one AR based on the NOAA definition.

Some regions are too small in areas or too weak in EUV flux to be detectable 
in the helioseismic far-side images, so it is meaningful to only count 
and compare the regions exceeding both a certain area in size and a certain value in EUV
flux density. Note that such small ARs may not cause a significant space 
weather concern, or cause a meaningful change in the large-scale magnetic 
field configuration. In this statistical study, we count the areas that have 
a logarithm value greater than 7.0 as the areas of the AR, and only those 
regions with a total area greater than 10 HIPs are counted as ARs in the 
following analysis. For the helioseismic images, we count the areas with a 
magnitude of phase shift greater than 0.06 radians, and again only those 
regions with a total area greater than 10 HIPs are counted as one AR in the 
statistical study. The selections of 7.0 as an EUV flux criterion, 0.06 as 
a helioseismic phase-shift criterion, and 10 HIPs as the minimum area for an 
AR are rather arbitrary, but as we can see in our later analysis, these selection
criteria can only affect the success rate for the lower end of the statistics,
but have little impact on large ARs.

We plot the EUV flux maps on top of the contemporaneous helioseismic images, and 
identify the regions that are counted as ARs in both types of maps based 
on the criteria given above. Understandably, helioseismic imaging cannot give 
a precise location for ARs due to its limit in spatial and temporal resolution, 
thus ARs that are identified as a positive match in EUV and helioseismic images 
may not entirely overlap each other. In practice, if an EUV AR overlaps 
with a helioseismic AR in 10 or more HIPs, we count them as a match. It is 
also found true that a large EUV AR may only correspond to a small helioseismic 
AR, or vice versa; we still count them as a positive match as long as they 
have 10 or more HIPs overlapping each other regardless of their respective
sizes.

\begin{figure}[!ht]
\epsscale{1.00}
\plotone{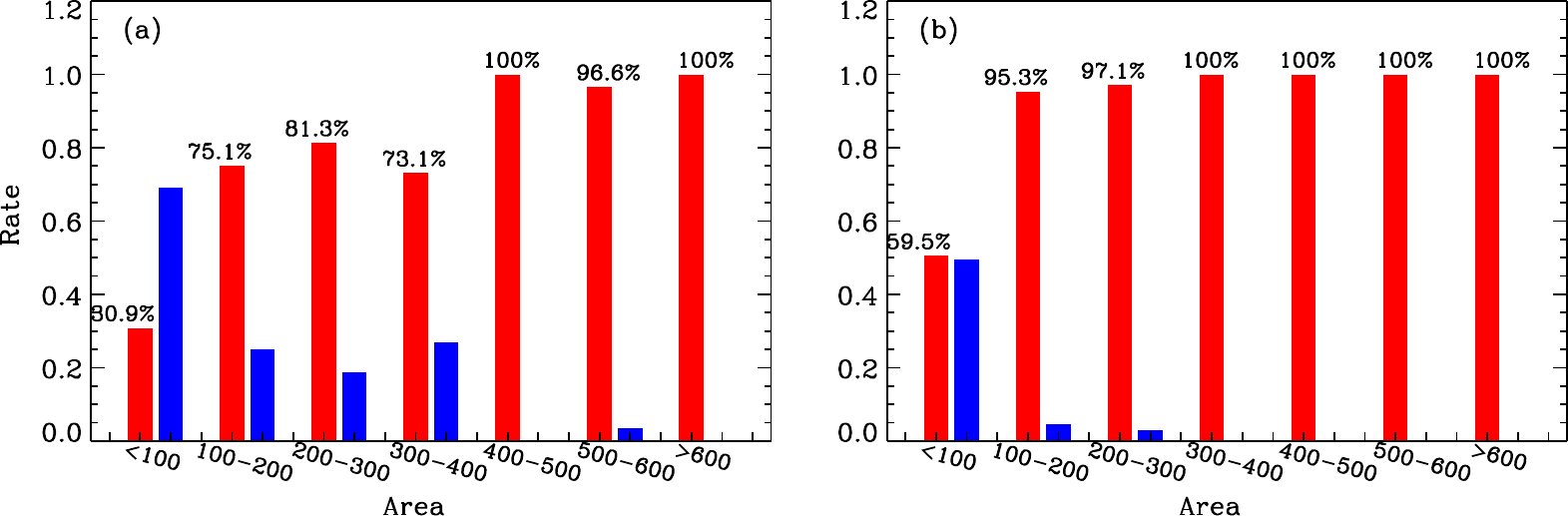}
\caption{(a) Percentage of EUV ARs that correspond to (red) or fail to correspond
to (blue) a helioseismic AR. (b) Percentage of helioseismic ARs that correspond
to (red) or fail to correspond to (blue) an EUV AR. The percentages are 
categorized based on the areas of EUV ARs (a) and helioseismic ARs (b), which
are plotted as horizontal axes with units of HIPs.} 
\label{f9}
\end{figure}

\subsection{Success Rate}
\label{sec53}

It is not surprising that the AR-detection success rate of helioseismic imaging 
depends on the AR sizes. We have identified a total of 1022 ARs in the 90-day 
EUV maps, and a total of 777 ARs in the helioseismic images during the same 
period. Note that many of these ARs are not independent ARs, because we do not 
follow ARs but treat every single far-side image as a random snapshot of the 
far side. 

To break down the numbers, for the EUV-identified ARs, of the 512 ARs with
a size between 10 and 100 HIPs, 158 (30.9\%) correspond to an AR detected in the 
helioseismic images; of the 287 EUV ARs with a size of 100 -- 200 HIPs, 
211 (75.1\%) correspond to an AR in the helioseismic images; of the 91 
EUV ARs with a size of 200 -- 300 HIPs, 74 (81.3\%) correspond to an AR 
in the helioseismic images; of the 52 EUV ARs with a size of 300 -- 400 HIPs, 
38 (73.1\%) correspond to an AR in the helioseismic images; of the 25 EUV 
ARs with a size of 400 -- 500 HIPs, all (100\%) of them correspond to an 
AR in the helioseismic images; of the 29 EUV ARs with a size of 500 -- 600 
HIPs, 28 (96.6\%) correspond to an AR in the helioseismic images; of the 
26 EUV ARs with a size greater than 600 HIPs, all (100\%) correspond to an 
AR in the helioseismic images. A histogram of this statistic is shown 
in Figure~\ref{f9}a. 

From the other perspective, of the 418 helioseismically-detected ARs that have
a size between 10 and 100 HIPs, 211 (50.5\%) correspond to an EUV-observed 
AR; of the 149 helioseismic ARs with a size of 100 -- 200 HIPs, 142 (95.3\%) 
correspond to an EUV AR; of the 68 helioseismic ARs with a size of 200 -- 300 
HIPs, 66 (97.1\%) correspond to an EUV AR; of the 59 helioseismic ARs 
with a size of 300 -- 400 HIPs, 27 helioseismic ARs with a size of 400 -- 500 
HIPs, 25 helioseismic ARs with a size of 400 -- 500 HIPs, and 30 helioseismic 
ARs with a size greater than 600 HIPs, all (100\%) of them correspond to
an EUV AR. A histogram of this statistic is shown in Figure~\ref{f9}b.

Overall, 85.7\% of all EUV ARs with a size larger than 200 HIPs are detected 
by the helioseismic imaging technique, and nearly 100\% of the ARs larger 
than 400 HIPs are detected helioseismically. Even better, almost all (97.3\%) 
of the helioseismic-identified ARs with an area greater than 100 HIPs, 
correspond to an observed EUV AR. This indicates that not all of the far-side ARs 
can be detected helioseismically, but the helioseismically detected regions, 
as long as their sizes are greater than a certain size, almost all correspond 
to a true far-side AR.

\section{Discussion and Conclusion}
\label{sec6}

Both the helioseismic holography and time--distance helioseismic methods for
imaging the Sun's far-side active regions have existed for many years. In this 
paper, building upon existing time--distance far-side imaging codes and 
expanding the number of measurement schemes from the previous 5 to the 
current 14 (see Table~\ref{tb1}), we have demonstrated that including more 
measurement schemes and using waves with a higher number of surface reflections 
substantially enhance the reliability and stability of far-side AR images.

Through comparing our new far-side images with the existing far-side images
made from the helioseismic holography technique, we find that the new method 
has a substantially better sensitivity in detecting ARs near both the 
far-side limbs. Particularly, a successful detection of ARs near the east 
limb, one or two days before their appearance on the near side of the Sun, 
is very useful for predicting the arrival of major ARs, hence improved space 
weather forecasting. We also find that our new method has an advantage over the 
holography method in earlier detection of newly emerged ARs on the far side. 
Because the emergence of a large AR can significantly alter the global-scale 
magnetic field configuration, the information on the new emergence is 
important to the global-scale modeling of the Sun's coronal magnetic-field 
structure and to modeling the solar wind. The helioseismic holography 
method uses a 5-day average to improve the stability of maps while our 
method uses 4-day averages, and this plays a role in the later 
detection of new emerging regions by the holography method. However, on the 
other hand, improving the stability of maps through overlapping a shorter 
period of data is itself an improvement. We also acknowledge that 
the holography method does a better job than our new method in ridding the 
systematic effects (positive patches) in high-latitude areas.

By comparing our new far-side images with \stereo/EUVI 304~\AA\ observations, 
we find that our new method is capable of detecting the majority of far-side 
ARs of medium and large size, with a good approximation in size, location, 
and strength; however, only the ARs larger than a certain size and stronger 
than a certain flux threshold are reliably detected by the helioseismic imaging 
method. Our statistics shows that about 85.7\% of ARs observed by \stereo\ 
with a size larger than 200 HIPs are detectable in the helioseismic images, 
while nearly all the helioseismically-detected ARs with an area larger 
than 100 HIPs correspond to an AR observed by \stereo. Therefore, every 
helioseismically-detected AR, as long as its size is greater than 100 HIPs, 
corresponds to an AR with very few number of false positive alarms. 
However, we need to caution that the area sizes given here are after the original 
\stereo\ 304~\AA\ images are convolved with a Gaussian function, hence 
substantially larger than the true sizes of those ARs. 

A comparison of helioseismic far-side images with the \stereo-observed 304~\AA\ 
images, and a comparison of the \sdo/HMI-observed magnetic-field maps and the 
\sdo/AIA-observed EUV images, together demonstrate that the helioseismic 
far-side images have a great potential to be calibrated into far-side maps 
of magnetic flux by using the 304~\AA\ images as a stepping-stone. If 
successful, it will help solar physicists better monitor the activity on
the Sun's far side, and better model the solar wind and magnetic field in
the corona.

Finally, it is worthwhile pointing out that the availability of our new 
time--distance helioseismic far-side imaging method complements the 
existing helioseismic holography method. A combination of the results from 
both methods will substantially boost the confidence of imaging the Sun's
far-side ARs, and therefore provide more reliable far-side information to the solar 
physics community.

\acknowledgments This work utilizes Dopplergrams and magnetograms observed 
by \sdo/HMI, and 304~\AA\ observations by \sdo/AIA and both \stereo/EUVI
instruments. The helioseismic holography far-side images are routinely 
produced and distributed by \sdo/JSOC. Both \sdo\ and \stereo\ are NASA 
missions. We thank Drs.~J.~Qiu and P.~Liewer for providing us the composite 
304~\AA\ maps made from \sdo/AIA and \stereo/EUVI observations. This work 
is partly supported by the NOAA grants WC-133R-16-CN-0116 and NA18NWS4680082.

\end{document}